\documentclass[oribibl]{llncs}
\usepackage[utf8]{inputenc}
\usepackage[tstt]{backnaur}
\usepackage{syntax}
\usepackage{graphicx} 
\usepackage{amsmath}
\usepackage{textcomp}
\usepackage{semantic}
\def\mspaceend{}%\raisebox{-6pt}{\strut}}
\def\mspacebegin{}%\raisebox{7pt}{\strut}}
\newcommand{\etal}{{et al.}}
\newcount\hour \newcount\minute %\newcount\second
\hour=\time \divide \hour by 60
\minute=\time %\divide \minute by 60
\loop \ifnum \minute > 59 \advance \minute by -60 \repeat
\def\currenttime{\the\hour:\the\minute}
\let\oldparagraph\paragraph
\def\paragraph#1{\oldparagraph{\textbf{\em #1}}}
\def\subsection#1{\paragraph{#1.}}
\newcommand{\tq}[0]{\mathtt{\times_q}}
\newcommand{\mq}[0]{\,\mathtt{-_q}\,}
\newcommand{\pq}[0]{\,\mathtt{+_q}\,}
\newcommand{\dq}[0]{\,\mathtt{/^q}\,}
\newcommand{\timesa}[0]{\mathtt{\times_i}}
\newcommand{\minusa}[0]{\mathtt{-_{\!i}}}
\newcommand{\plusa}[0]{\mathtt{\,+_{\!i}\,}}
\newcommand{\diva}[0]{\,\mathtt{\%_i}\,} 
\newcommand{\eqa}[0]{\mathtt{\,=_i\,}}
\newcommand{\leqa}[0]{\mathtt{\,\leq_i}}

\newcommand{\sumq}[2]{\mathtt{sum(}#1 \mathtt{\,;\,} #2 \mathtt{)}}
\newcommand{\prodq}[3]{\mathtt{prod(}#1 \mathtt{\,;\;} #2 \mathtt{\,;\;} #3 \mathtt{)}}

\newcommand{\f}[1]{\mathtt{f(}#1\mathtt{)}\,}
\newcommand{\Pt}[1]{\,\mathtt{P(}#1\mathtt{)}}

\newcommand{\Ct}[1]{\mathtt{c(}#1\mathtt{)}}
\newcommand{\ldotss}{...}
\newcommand{\eqdef}{\mathrel{\overset{\makebox[0pt]{\mbox{\normalfont\tiny\sffamily def}}}{=}}}
\newcommand{\x}[1]{\mathtt{#1}}

\def\Pup{\overline{P}} %

\def\enddot{\hspace*{1mm}\mbox{.}}

\DeclareMathOperator{\Var}{\mathcal{V}\mathit{ar}} % Var

\newcommand{\Const}{\mathtt{n}}
\def\comment#1{\marginpar{\hbox to 0pt{\vbox to 0pt{%
  \framebox[40mm]{\hskip 1mm\parbox{37mm}{#1}\hskip 1mm}%
  \vss}\hss}%
}}
\title{Probabilistic Resource Analysis by Program Transformation} %MHT 
\institute{Computer Science, Roskilde University\\
Roskilde, Denmark\\
\email{majaht@ruc.dk}, \email{madsr@ruc.dk} \ourthanks
}
\author{Maja H. Kirkeby \and{} Mads Rosendahl}
\def\ourthanks{\footnote{%
 The research leading to these results has received funding from
 the European Union Seventh Framework Programme (FP7/2007-2013)
 under grant agreement no 318337,
 ENTRA - Whole-Systems Energy Transparency.}}

\begin{document}

\maketitle 
\begin{abstract}
The aim of a probabilistic resource analysis is to derive a probability distribution of possible resource usage for a program from a probability distribution of its input. 
We present an automated multi-phase rewriting based method to analyze programs  written in a subset of C. 
It generates a probability distribution of the resource usage as a possibly uncomputable expression and then transforms it into a closed form expression using over-approximations. 
We present the technique, outline the implementation and show results from experiments with the system.
%
%\comment{\today: \currenttime}
\end{abstract}
\def\comment#1{}

%-----------------------------------------------------
\section{Introduction}
The main contribution in this paper is to present a technique for probabilistic resource analysis where the analysis is seen as a program-to-program translation. 
This means that the transformation to closed form is a source code program transformation problem and not specific to the analysis. Any necessary approximations in the analysis are performed at the source code level.
The technique also makes it possible to 
balance the precision of the analysis against the brevity of the result.

Many optimizations for increased energy efficiency require probabilistic and average-case analysis as part of the transformations.
%When analyzing energy consumption, probability distributions may provide more useful information than boundaries. 
Wierman et al. state that \textit{``global energy consumption is affected by the average-case, rather than the worst case``} \cite{conf:allerton:Wierman2008}. 
Also in scheduling \textit{``an accurate measurement of a task's average-case execution time can assist in the calculation of more appropriate deadlines''} \cite{jour:cc:Guo2007}. 
For a subset of programs a precise average-case execution time can be found using static analysis \cite{journals/tcs/FlajoletSZ91,books/daglib/0020847,thesis:cork:gao2013}. 
Applications of such analysis may be in improving scheduling of operations 
or in temperature management. %, where worst-case bounds are important \cite{conf/rtas/SchorBYT12}. 
Because the analysis returns a distribution, it can be used to calculate the probability of energy consumptions above a certain limit, and thereby indicate the risk of over-heating. 

The central idea in this paper is to use probabilistic output analysis in 
combination with a preprocessing phase that instruments programs with resource usage.
We translate programs into an intermediate language 
program that computes the probability distribution of resource usage. 
This program is then analyzed, transformed, and approximated with the aim of obtaing a closed form expression. 
It is an alternative to deriving cost relations directly from the program \cite{Debray94}
%\cite{journals/entcs/AlbertAGP09,conf/iclp/Lopez-GarciaDB10}
or expressing costs as abstract values in a semantics for the language.

As with automatic complexity analysis, the aim of probabilistic resource analysis is to express the result as a parameterized expression. 
The time complexity of a program should be expressed as a closed form expression in the input size, and for probabilistic resource analysis, the aim is to express the probability of resource usage of the program parameterized by input size or range.
If input values are not independent, we can specify a joint distribution for the values. 
Values do not have to be restricted to a finite range but for infinite ranges the distribution would converge to zero towards the limit.

The current work extends our previous work on probabilistic analysis 
\cite{conf:qapl:Rosendahl2015} in three ways. 
We show how to use a preprocessing phase to instrument programs with resource usage such that the resource analysis can be expressed as an analysis of the possible output of a program. 
The resource analysis can handle an extended class of programs with structured data as long as the program flow does not depend on the probabilistic data in composite data structures. Finally, we present an implementation of the analysis in the Ciao language \cite{bueno1997ciao} which uses algebraic reductions in the Mathematica system
\cite{wolfram2000mathematica}.

The focus in this paper is on using fairly simple local resource measures where we count core operations on data. Since the instrumentation is done at the source code level, we can use flow information so that the local costs can depend on actual data to operations and which operations are executed before and after. This is normally not relevant for time complexity but does play an important role for energy consumption analysis
\cite{kerrison2013energy,tiwari1994power}. 

%-----------------------------------------------------
\section{Probability distributions in static analysis}

%Probabilistic output analysis analyze a program \texttt{f}: $Int^{*} \rightarrow Int$ and a discrete input probability distribution \texttt{P}: $Int^{*} \rightarrow Real$ which assigns a probability between 0 and 1 to each input value, and return a discrete output probability distribution \texttt{P$_{f}$}, or sometimes named output probability function.

In our approach to probabilistic analysis, the result of an analysis is an approximation of a probability distribution. We will here present the concepts and notation we will use in the rest of the paper.
%
%We consider the input to a program as a discrete random variable and the input probability distribution is then a probability measure that to an event of input having a given value assigns a value between 0 and 1. 
%
A probability distribution is also often referred to as the \emph{probability mass function} in the discrete case, and in the continuous case, it is a 
\emph{probability density functions}. 
%
%We will use the phrase \emph{probability function} to denote mappings from single values (input or output) to a probability or number between 0 and 1, and we 
We will use an upper case $P$ letter to denote a probability distribution.

\begin{definition}[input probability]\label{def:inputDist}
For a countable set $X$ an input probability distribution is a mapping $P_X:X\rightarrow \{r\in\bbbr\mid 0\leq r\leq 1\}$, where
\begin{align*}
 \sum_{x\in X} P_X(x) = 1  
% &\qquad\qquad\textrm{ , where }\forall\bar{x}\in X.0\leq P_x(\bar{x})\leq 1
\end{align*}
\end{definition}
We define the output probability distribution for a program {\tt p} in a forward manner. 
It is the {\em weight} or sum of all probabilities of input values where the program returns the desired value $z$ as output.

\begin{definition}[output probability]\label{def:outputDist}
 Given a program, $\mathtt{p} : X \rightarrow Z$ and a probability distribution for the input,  $P_{X}$, the output probability distribution, $P_{\mathtt{p}}(z)$, is defined as:
 \begin{align*}\mspacebegin
  P_{\mathtt{p}}(z) & = \sum_{x \in X \land \mathtt{p}(x) = z} P_{X}(x) 
 \end{align*}
\end{definition}
Note that Kozen also uses a similar forward definition 
\cite{journals/jcss/Kozen81}, 
%\cite{conf/focs/Kozen79},
whereas Monniaux constructs the inverse mapping from output to input for each program statement and expresses the relationship in a backwards style  \cite{conf/sas/Monniaux00}. 
\begin{lemma}\label{lem:sumOutBound}
The output probability distribution, $P_{\mathtt{p}}(z)$, 
satisfies
\begin{align*}\mspacebegin
 0 \leq \sum_{z} P_{\mathtt{p}}(z) \leq 1
\mspaceend
 \end{align*}
\end{lemma}
The program may not terminate for all input, and this means that the sum may be less than one. 
If we expand the domain $Z$ with an element to denote non-termination, $Z_\bot$, the total sum of the output distribution $P_{\mathtt{p}}(z)$ would be 1.

In our static analysis, we will use approximations to obtain safe and simplified results.
Various approaches to approximations of probability distributions have been proposed and can be interpreted as {\em imprecise probabilities}
\cite{conf:vstte:AdjeB2014,tech:compegne:Ferson2014,conf:isip:Destercke2009}. 
Dempster-Shafer structures \cite{conf:rbes:Gordon1984,conf/uai/Bauer96} 
and P-boxes \cite{tech:sand:Ferson2002} can be used to capture and propagate uncertainties of probability distributions. 
There are several results on extending arithmetic operations to probability distributions for both known dependencies between random variables and when the dependency is unknown or only partially known
\cite{journals/rc/BerleantC98,journals/computing/BouissouGGP12,conf:rta:Kay2007,thesis:Stellenbosch:Uwimbabazi2013,conf:hdrums:Wilson2000}. 
Algorithms for lifting basic operations on numbers to basic operations on probability distributions can be used as abstractions in static analysis based on abstract interpretation.
Our approach uses the P-boxes as bounds of probability distributions.
P-boxes are normally expressed in terms of the cumulative probability distribution but we will here use the probability mass function. We do not, however, use the various basic operations on P-boxes, but apply approximations to a probability program such that it forms a P-box.

\begin{definition}[over-approximation]\label{def:approximationP}
For a distribution $P_{\mathtt{p}}$ an over-approxi\-mation 
 ($ \Pup{}_{\mathtt{p}}$)
of the distribution satisfies the condition:
 \begin{align*}\mspacebegin
  & \Pup{}_{\mathtt{p}} :  \forall z . 
    P_{\mathtt{p}}(z) \leq  \Pup{}_{\mathtt{p}}(z) \leq 1
\enddot
\mspaceend      
 \end{align*}
\end{definition}
The aim of the probabilistic resource analysis is to derive an approximation
$ \Pup{}_{\mathtt{p}}$ as tight as possible.

The over-approximation of the probability distribution can be used to derive 
lower and upper bounds of the expected value and will thus approximate the expected value as an interval
\cite{conf:qapl:Rosendahl2015}.

%---------------------------------------------------------------------
\section{Architecture of the transformation system}

The system contains five main phases. 
The input to the system is a program in a small subset of C with annotations of which part we want to analyze. It could be the whole program but can also be a specific subroutine which is called repeatedly with varying arguments according to some input distribution.

The first phase will instrument the program with resource measuring operations. The instrumented program will perform the same operations as the original program in addition to recording and printing resource usage information. This program can still be compiled and run, and it will also produce the same results as the original program.

The second phase translates the program into an intermediate language for further analysis. We use a small first-order functional language for the analysis process. The translation has two core elements. We slice \cite{weiser1981program} the program with respect to the resource measuring operations and transform loops into primitive recursion in the intermediate language. The transformed program can still be executed and will produce the same resource usage information as the instrumented program. Since the instrumentation is done before the translation to intermediate language any interpretation overhead or speed-up due to slicing does not influence the result \cite{conf/fpca/Rosendahl89}.

In the third phase, we construct a probability output program that computes the probability output function. In this case, it is a probability distribution of possible resource usages of the original program. This program can also run but will often be extremely inefficient since it will merge information for all possible input to the original program.

The fourth phase transforms the probability program into 
a large expression without further function calls. Recursive calls are removed using summations and the transformed program
computes the same result as the program did before this phase.

In the final phase the probability function is transformed into closed form using symbolic summation and over-approximation. In this phase we exploit the Mathematica system \cite{wolfram2000mathematica}. 
The final probability program computes the same result or an over-approximation of the function produced in the fourth phase.

%-----------------------------------------------------
\section{Instrumenting programs for resource analysis}

The input to the analysis is a program in a subset of C.
In the next section we define the intermediate language for further analysis and it is the restrictions on the intermediate language that limits the source programs we can analyze with our system. The source program may contain integer variable and arrays, usual loop constructs and non-recursive function calls.
The program should be annotated with specification on which part of the program to analyze. The following is an example of such a program.

\begin{quote}\small\begin{verbatim}
// Toanalyze: multa(_,_,_,N)
void multa(int a1[MX],int a2[MX],int a3[MX],int n){
  int i1,i2,i3,d;
  for(i1 = 0; i1 < n; i1++) {
    for(i2 = 0; i2 < n; i2++) {
      d = 0;
      for(i3 = 0; i3 < n; i3++) {
         d = d + a1[i1*n+i3]*a2[i3*n+i2];
      } 
      a3[i1*n+i2] = d;
    }
  }
}
\end{verbatim}\end{quote}
This example program describes a matrix multiplication for which we would 
like to analyze the probability distribution for the number of steps when 
parameterized with the size (\texttt{N}) of the matrices.

\paragraph{Instrumentation.}
The program is then instrumented with resource usage information and translated into an intermediate language for further analysis. 
The instrumented program is also a valid program in the source language and can be executed with the same results as the original program. It will, however, also collect resource usage information.

In our example, we instrument the program with step counting information where we count 
the number of assignment statement being executed. This is done by inserting a variable into the program and incrementing it once for each assignment statement.

\begin{quote}\small\begin{verbatim}
int multa(int a1[MX],int a2[MX],int a3[MX],int n){
  int i1,i2,i3,d;
  int step; step=0;
  for(i1 = 0; i1 < n; i1++) {
    for(i2 = 0; i2 < n; i2++) {
      d = 0; step++;
      for(i3 = 0; i3 < n; i3++) {
         d = d + a1[i1*n+i3]*a2[i3*n+i2]; step++;
      } 
      a3[i1*n+i2] = d; step++;
    }
  }
  return step;
}
\end{verbatim}\end{quote}
The outer loop does not update the step counter, whereas the first inner loop 
updates it twice per iteration and the innermost loop updates it once per loop 
iteration.

\paragraph{Slicing.}
The second phase will slice the program with respect to resource usage and translate the program into the intermediate language of first-order functions that we will use in the subsequent stages. Loops in the program are translated into primitive recursion.

\begin{quote}\small\begin{verbatim}
for3(i3, step, n) =
  if(i3 = n) then step else for3(i3 + 1,step+1,n)

for2(i2, step, n) =  
  if(i2 = n) then step else for2(i2 + 1,for3(0,step+2,n),n)

for1(i1, step, n) =
  if(i1 = n) then step else for1(i1 + 1,for2(0,step,n),n)

tmulta(n)= for1(0,step,n)
\end{verbatim}\end{quote}
Each function in the recursive program corresponds to a for loop with their 
related step-updates. The step counter is given as input argument to the next 
function in a continuation-passing style.

\paragraph{Intermediate language.}

An intermediate program, \texttt{Prg}, consists of integer functions, \texttt{f$_i$}: $Int^{*} \rightarrow Int$, as given by the abstract syntax given in Figure \ref{fig:syntax-f}. In the following, we relax the restrictions on function and parameter names.

\begin{figure}[h!]
\begin{bnf*} 
\bnfts{f}_i\bnfts{(}x_1\bnfts{,} \bnfsk \bnfts{,}x_n\bnfts{)} & \eqdef &  \bnfpn{exp} \\ % \quad \textrm{:} int^* \rightarrow int\\
\bnfprod{aexp}
{\bnfts{x}_i \bnfor 
\bnfts{c} \bnfor 
\bnfpn{aexp} \; \bnfts{+}_{\bnfts{i}}\; \bnfpn{aexp} \bnfor
\bnfpn{aexp} \; \bnfts{-}_{\bnfts{i}}\; \bnfpn{aexp} \bnfor   \\ &&
\bnfpn{aexp} \; \bnfts{$\times$}_{\bnfts{i}}\; \bnfpn{aexp} \bnfor 
\bnfpn{aexp} \; \bnfts{div}_{\bnfts{i}} \; \bnfpn{aexp}}
\\
\bnfprod{bexp}{
\bnfpn{aexp} \;\bnfts{=}_{\bnfts{i}}\; \bnfpn{aexp} \bnfor 
\bnfpn{aexp} \;\bnfts{<}_{\bnfts{i}}\; \bnfpn{aexp} \bnfor 
\bnfpn{aexp} \;\bnfts{$\leq$}_{\bnfts{i}}\; \bnfpn{aexp} \bnfor \\&&
%\bnfpn{bexp} \;op_b\; \bnfpn{bexp} \bnfor 
\bnfts{true} \bnfor 
\bnfts{false} \bnfor 
\bnfts{not(}\bnfpn{bexp}\bnfts{)}
}\\
\bnfprod{exp}
{\bnfpn{aexp} \bnfor
\bnfts{f$_i$(}\bnfpn{exp$_1$}\bnfts{, ...,} \bnfpn{exp$_n$}\bnfts{)}   \bnfor \\&&
\bnfts{if } \bnfpn{bexp} \bnfts{ then } \bnfpn{exp} \bnfts{ else } \bnfpn{exp}} 
\end{bnf*} 

\caption{The abstract syntax describing the intermediate programs.} 
\label{fig:syntax-f}
\end{figure}

\begin{definition}\label{def:wellformed}
 A program is \emph{well-formed} if it follows the abstract syntax and it contains a finite number of function definitions, that each is of one of the following forms and can internally be enumerated with a natural number such that:
 \[
 \begin{array}{l}
     \mathtt{f_{i}}(x_1, \ldots, x_n) \eqdef \mathtt{if\;} {b} \mathtt{\;then\;} e_0 \mathtt{\;else\;} \mathtt{f_{i}}(e_{1}, \ldots, e_{n}) \\
    \text{ where } \mathtt{f_{i}} \text{ is simple, } e_0 \text{ only contains calls to functions }\mathtt{f_{j}}\text{ where }j<i. \\ \\
   % \mathtt{f_{i}}(x_1, \ldots, x_n) \eqdef \mathtt{if\;} {b} \mathtt{\;then\;} e_0 \mathtt{\;else\;} e_1 \\
   % \text{ where }k < i \text{ and }  e_0, e_{1} \text{ only contain calls to functions }\mathtt{f_{j}}\text{ where }j<i. \\ \\
    \mathtt{f_{i}}(x_1, \ldots, x_n) \eqdef e \\ \text{ where } e \text{ only contain calls to functions }\mathtt{f_{j}}\text{ where }j<i.\\
  \end{array}
 \]
\end{definition} 
The enumeration prevents mutual recursion and ensures that non-recursive calls cannot create an infinite call-chain. 

\section{Probabilistic output analysis} 
The analysis is applied to
the intermediate program and an input probability program in 
the intermediate language. The output is a new program that can be described by a subset of the intermediate language; this will be clarified later in the definition of pure and closed form programs.
The analysis consists of three phases: 
\begin{description}
 \item{Create}, where the probability program describing the output distribution is created as a possibly uncomputable expression. 
 \item{Separate}, where we remove all calls from the probability program.
 \item{Simplify}, where we transform the program into closed form using safe over-approximations when necessary.
\end{description}
The analysis is constructed as three sets of transformation rules, one for each of the three phases. All transformations are syntax directed, and a strategy is to apply them in a depth-first manner. 
The program output analysis is implemented in Ciao and integrates with Mathematica in the third phase to reduce expressions.

In the following we use $\Var(e)$ to represent the set of variables occurring in expression $e$, and $\f{\x{x}_1, \ldotss, \x{x}_n} \eqdef e $ to represent the function $\x{f}$ is defined in the input program. 
Some side conditions are explained in an informal way, as in
``$\f{\x{x}_1, \ldotss, \x{x}_n} \eqdef e \text{, where } e \text{ is non-recursive}$''.
\[
\inference[name]
{\text{precondition}_1 & \ldotss & \text{precondition}_n}
{\text{original term} -> \text{rewritten term}}
\]
The preconditions are evaluated from left to right, and if all succeeds, we can use the transformation.%, and preconditions may use the $\Var$ or require a rewriting of subterm. 
When substituting a variable $\x{x}$ to an expression $e$, we denote it $[\x{x}/e]$. %We may use the standard concept of context when describing a check for a certain subterm in a term; given a term $t$ and some term $e$, we may write $\exists s . t = s[e]$ for the check of whether $e$ exists as a subterm in $s$. The context $s$ may then be used for replacing the subterm $e$ with a new subterm, written as $s[e']$.

In the following we will begin by extending the intermediate language presented in Figure \ref{fig:syntax-f} such that it can express probabilities, and afterwards describe the transformation rules for each phase.  

% ----------------------------------------------------------------------
\subsection{The intermediate language}

The intermediate language is, as previously mentioned, a first-order functional language. A probability program can 
be evaluated at any stage through the transformation process. 
%always be evaluated, and since each rule transforms between probability programs it is always possible to evaluate the current state of the program. 
%We assume no nested calls in probability functions.

We extend the abstract syntax given in Figure \ref{fig:syntax-f} such that it can easily describe probability distributions. We introduce probability functions, \texttt{P}: $Int^{*} \rightarrow Real$, which follows the expanded syntax given in Figure \ref{fig:syntax-p}. The dots indicate the syntax described in Figure \ref{fig:syntax-f}. Again, $\bnfpn{aexp}$ and $\bnfpn{exp}$ are of type integer, $\bnfpn{bexp}$ is boolean, and the new $\bnfpn{qexp}$ is a real. In $\bnfpn{qexp}$ the method \texttt{i2r} type casts an integer expression to a real. 
We introduce \texttt{c}, \texttt{sum}, \texttt{prod} and \texttt{argDev} functions. \texttt{c} evaluates to either 1, if its boolean expression evaluates to true, or 0 when it evaluates to false. Evaluating \texttt{sum} instantiates the variable with all possible values and sum all the results of the evaluation the $\bnfpn{qexp}$. \texttt{prod} instantiates its variable with all values for which the first  $\bnfpn{qexp}$ evaluates to 1, and then it multiply all the results from evaluating the second $\bnfpn{qexp}$. 
%The \texttt{prod} function is used for describing a constraint (second $\bnfpn{qexp}$) that must be true for a range of values (first $\bnfpn{qexp}$). 
The last expression introduced is $\bnfts{argDev}$ which describes the development of the variable $\bnfts{x}_i$ as a function of the number of updates, $\bnfts{x}_j$. The expression $\bnfpn{exp}$ computes the development of $\bnfts{x}_i$ for one incrementation of $\bnfts{x}_j$ (e.g. the argument $\x{x}_i$ in a function $\x{f(x}_i\x{)}$ with a recursive call $\x{f(x}_i\minusa\x{2)}$ has a argument development  $\bnfts{argDev(x}_i\bnfts{,x}_i\x{-2,x}_j\x{)}$). 

\begin{figure}[h!]
\begin{bnf*} 
\bnfts{f}_i\bnfts{(}x_1\bnfts{,} \bnfsk \bnfts{,}x_n\bnfts{)} & \eqdef &  \bnfpn{exp} \\ % \quad \textrm{:} int^* \rightarrow int\\
\bnfprod{aexp}
{\ldots \bnfor
\bnfts{min(} \bnfpn{aexp} \bnfts{,} \bnfpn{aexp} \bnfts{)} \bnfor 
\bnfts{max(} \bnfpn{aexp} \bnfts{,} \bnfpn{aexp} \bnfts{)}} \\ 
\bnfprod{bexp}{ \ldots \bnfor
\bnfpn{aexp} \;\bnfts{=}_{\bnfts{i}}\; \bnfpn{exp} 
}\\
\bnfprod{exp}{ \ldots \bnfor \bnfts{argDev(} \bnfts{x}_i , \bnfpn{exp} ,\bnfts{x}_j \bnfts{)}} \\
\bnfts{P}_i\bnfts{(}x_1\bnfts{,} \bnfsk \bnfts{,}x_n\bnfts{)} & \eqdef&  \bnfpn{qexp} \\
\bnfprod{qexp}{
\bnfts{i2r(}\bnfpn{aexp}\bnfts{)} \bnfor 
\bnfts{c(} \bnfpn{bexp} \bnfts{)} \bnfor
\bnfpn{qexp} \;op_q\; \bnfpn{qexp} \bnfor \\&&
\bnfts{sum(} \bnfts{x}_i , \bnfpn{qexp} \bnfts{)} \bnfor
\bnfts{prod(} \bnfts{x}_i , \bnfpn{qexp} , \bnfpn{qexp} \bnfts{)} \bnfor \\&&
\bnfts{P$_i$(}\bnfpn{aexp$_1$}\bnfts{, ...,} \bnfpn{aexp$_n$}\bnfts{)}
}\\
op_q & = & \bnfts{+}_{\bnfts{q}} \bnfor 
\bnfts{-}_{\bnfts{q}} \bnfor 
\bnfts{$\times$}_{\bnfts{q}} \bnfor 
\bnfts{/}^{\bnfts{q}} 
\end{bnf*} 
\caption{The expanded abstract syntax describing probability programs.} 
\label{fig:syntax-p}
\end{figure}

 A program that computes a probability distribution is referred to as a probability program. 
\begin{definition}
A probability program that has no  if-expressions no function calls is 
\emph{pure} and a pure probability program without any \texttt{sum} and 
\texttt{prod} is in \emph{closed form}. 
\end{definition}
A program is \emph{pure} after it is transformed in the separation phase and is pure and in \emph{closed form} after the simplification phase.

\subsection{The create phase}
This phase has only one rule 
which creates a program that computes a probability distribution from the intermediate program and input distributions.
\[
\begin{array}{@{}l@{}}
\inference[\textsf{create}]
{\f{\x{u}_1, \ldotss, \x{u}_n} \eqdef e & %\mathtt{f}: \mathbb{I}^{*} -> \mathbb{I}^{*} & 
\Pt{\x{v}_1, \ldotss, \x{v}_n} \eqdef e_p %& %\mathtt{P}:\, \mathbb{I}^{*} -> \mathbb{Z} & 
%\x{u}_1, \ldotss, \x{u}_n \text{ are fresh variables}
}
{\mathtt{P}_{\mathtt{f}}(\x{z}) \eqdef \sumq{\x{x}_1}{\ldotss \sumq{\x{x}_n}{\Ct{\x{z} \eqa \f{\x{x}_1, \ldotss, \x{x}_n}} \tq \Pt{\x{x}_1, \ldotss, \x{x}_n}}}
}
\end{array}
\]
{We use the create rule to make a new probability function describing the probability distribution for the integer function we are interested in.}

\subsection{The separate phase}
In this phase function calls are removed by repeatedly exposing calls and replacing them. 
Non-recursive function calls are unfolded using their definitions. Function calls can occur inside if-expressions or as nested calls; these are extracted and handled one at a time.
{\footnotesize \[
\begin{array}{@{}l@{}}
\inference[f-simple]
{\f{\x{y}_1, \ldotss, \x{y}_n} \eqdef e &\text{, where } e \text{ is non-recursive} & \x{x}_1, \ldotss, \x{x}_n \in \Var}
{\Ct{\x{z}  \eqa \f{\x{x}_1, \ldotss, \x{x}_n}\!} -> \Ct{\x{z} \eqa e[\x{y}_1/\x{x}_1, \ldotss, \x{y}_n/\x{x}_n]}}\\\\
%%%%%%%%%%%%%%%%%%%%%%%%%%%%%%%%%%%%%%
\inference[rem-P]
{\Pt{x_1, \ldotss, x_n} \eqdef e }
{\Pt{e_1, \ldotss, e_n} -> e[x_1 / e_1, \ldotss, x_n / e_n]}\\ \\
\inference[rem-if]
{}
{
\Ct{\x{z} \eqa \mathtt{if\,} b \mathtt{\,then\,} e_0 \mathtt{\,else\,} e_1 }-> %\\
\left( \Ct{b} \tq  \Ct{\x{z} \eqa\! e_0} \pq \Ct{\mathtt{not}(b)} \tq  \Ct{\x{z} \eqa\! e_1} \right)
}
\\ \\
 %%%%%%%%%%%%%%%%%%%%%%%%%%%%%%%%%%%%% 
\inference[no-nest(f)]
{\{e_1, \ldotss, e_n\} \not\subseteq \Var}{
\begin{array}{@{}l@{}}
  \Ct{\x{z} \eqa \f{e_1, \ldotss, e_n}} 
     -> \\ 
  \sumq{\mathtt{u}_1}{\ldotss \sumq{\mathtt{u}_n}{ 
	      \Ct{\x{z} \eqa \f{\x{u}_1, \ldotss, \x{u}_n}} \tq \Ct{\x{u}_1 \eqa e_1} \tq \ldotss \tq \Ct{\x{u}_n \eqa e_n}
    \,}}\\
 \end{array}
}
\end{array}
\]}
We replace calls to recursive functions by a summation over the number of recursions using argument development constructors to describe the value of each argument as a function of the index of the summation. 
This way of defining argument development has similarities with size change functions derived using recurrence equations. 
Argument development functions do not depend on the base-case unlike size-change functions \cite{zuleger2011bound}.
The summation also contains a product which ensures that the condition evaluates to false for argument values less than the current value of the index of summation. 
When the expression in a product contains only c-constructors, then the product is evaluated to 1 if either the range is empty or the expression is evaluated to true for the full range. 
The following rewrite rules are all that is needed for transforming probability programs into pure probability programs.
{\footnotesize
 \[
\begin{array}{@{}l@{}}
  \inference[f-rec]
{
\mathtt{f}(\x{y}_1, \ldotss, \x{y}_n) \eqdef \mathtt{if\,\,} b \mathtt{\,\,then\,\,} e_0 \mathtt{\,\,else\,\,} \mathtt{f}(e_1, \ldotss, e_n) &  \x{x}_1, \ldotss \x{x}_n \in Vars\\ 
 \sigma_{y/i} = [\x{y}_1/\x{i}_1, \ldotss, \x{y}_n/\x{i}_n]  \sigma_{y/x} = [\x{y}_1/\x{x}_1, \ldotss, \x{y}_n/\x{x}_n]  \sigma_{y/j} = [\x{y}_1/\x{j}_1, \ldotss, \x{y}_n/\x{j}_n]
}
{
\begin{array}{@{}c@{}}
\Ct{\x{z} \eqa \mathtt{f}(\x{x}_1, \ldotss, \x{x}_n)} -> \qquad \qquad \qquad \qquad \qquad \qquad \qquad \qquad \qquad \qquad \qquad \qquad \\
\begin{array}{@{}l@{}}
\sumq{\x{i}}{
  \Ct{0 \leqa \x{i}} \tq \\
  \;\; \sumq{\x{i}_1}{\ldotss \sumq{\x{i}_n}{ \Ct{\sigma_{y/i}(b)} \tq  
   \Ct{\x{i}_1 \eqa \x{argDev(}\x{x}_1,\sigma_{y/x}(e_1),\x{i}\x{)}} \tq \\
  \;\; \qquad
       \Ct{\x{z} \eqa \sigma_{y/i}(e_0)} \tq
     \ldotss \tq \Ct{\x{i}_n \eqa \x{argDev(}\x{x}_n,\sigma_{y/x}(e_n),\x{i}\x{)}}
  \; }\ldotss} \tq \\
   \;\; \prodq{\x{j}}{ \Ct{0 \leqa \x{j}} \tq \Ct{\x{j} \leqa \x{i} \minusa 1}}{ \\
   \;\;
  \sumq{\x{j}_1}{\ldotss \sumq{\x{j}_n}{ \Ct{\mathtt{not}(\sigma_{y/j}(b))} \tq \\
  \;\; \qquad
   \Ct{\x{j}_1 \eqa \x{argDev(}\x{x}_1,\sigma_{y/x}(e_1),\x{j}\x{)}} \tq \ldotss \tq \Ct{\x{j}_n \eqa \x{argDev(}\x{x}_n,\sigma_{y/x}(e_n),\x{j}\x{)}} \\ 
   \;\;
  }\ldotss}\,}}
\end{array}
\end{array}
  }
 \end{array}
\]}
The argument development expression may contain function calls as well, and these are extracted equivalently to nested functions.
{\footnotesize
 \[
\begin{array}{@{}l@{}}
\inference[no-nest(argDev)]
{ }{
\begin{array}{@{}l@{}}
  \Ct{\x{z} \eqa \x{argDev(}\x{x}, \f{e_1, \ldotss, e_n},\x{i)}} 
     -> \\ 
  \sumq{\mathtt{u}}{ 
	      \Ct{\x{z} \eqa \x{argDev(}\x{x}, \f{e_1, \ldotss, e_n},\x{i)}} \tq \Ct{\x{u} \eqa \f{e_1, \ldotss, e_n}\!}
    \,}
\end{array}}
\end{array}
\]}
After applying these rules until they cannot be applied anymore, the probability program has been transformed to pure form.

\subsection{The simplification phase}
We have presented the rules for obtaining a pure probability program, and in this section we outline the rules used to reach closed form. 
A pure probability function consists of a series of nested summations multiplied with an expression (e.g. input probability). 
The rules are applied in no particular order and the phase ends when no more rules can be applied. 
In this phase we integrate with Mathematica. A call to Mathematica is denoted $\x{mm \!\!:\!\!Function(}Arg\x{)}=Answer$, where \texttt{Function} denotes the actual function called in Mathematica (e.g. $\x{mm \!\!:\!\!Expand}$ calls Mathematica's \texttt{Expand} function).
The translation between the intermediate language and Mathematica's representation
will not be discussed further here. 

The rules can be grouped by their functionality: preparing expressions, removal of summations and removal of products. 
The latter are currently the only rules containing over-approximations.

\subsubsection*{Preparing} 
expressions for removal of either summations or products involve moving expressions 
that do not depend on the index of summation outside the summation, dividing 
summations of additions into simpler ones, reducing expressions, dividing summations in ranges, and remove argument development constructors. Please notice that \textsf{div-sum(x$\leq$)} has an equivalent rule for upper bounds.
 \[
\begin{array}{l}
\inference[move-c]
{\mathtt{x} \notin \mathcal{V}ar(e_1)}
{\sumq{\mathtt{x}}{e_1 \tq e_2} -> 
 e_1 \tq \sumq{\mathtt{x}}{e_2}} 
\\[15pt]
\inference[div-sum(+)]
{\mathtt{x} \in \mathcal{V}ar(e_1) & \mathtt{x} \in \mathcal{V}ar(e_2)}
{\sumq{\mathtt{x}}{e_1 \pq e_2} -> 
 \sumq{\mathtt{x}}{e_1} \pq \sumq{\mathtt{x}}{e_2}}
\\[15pt]
\inference[div-sum(x$\leq$)]
{\mathtt{x} \notin \mathcal{V}ar(e_1,e_2) & \mathtt{x} \in \mathcal{V}ar(e_2)}
{
  \begin{array}{l}
  \sumq{\mathtt{x}}{\Ct{\x{x} \leqa e_1} \tq \Ct{\x{x} \leqa e_2} \tq e_3} -> \\
 \Ct{e_1 \leqa e_2} \tq \sumq{\mathtt{x}}{\Ct{\x{x} \leqa e_1} \tq e_3} \pq \\
 \Ct{e_2 \leqa e_1 \minusa 1} \tq \sumq{\mathtt{x}}{\Ct{\x{x} \leqa e_2} \tq e_3}
 \end{array}
 }
\\[15pt]
\inference[rem(argDev)]
{ \x{c} \in \Const}{
\begin{array}{@{}l@{}}
  \Ct{\x{z} \eqa \x{argDev(}\x{x},\x{x} \plusa \x{c},\x{i)}}
  ->
  \Ct{\x{z} \eqa \x{x} \plusa \x{c} \timesa \x{i}}   
\end{array}
}
\\[15pt] 
\inference[reduceAexp]{\x{mm \!\!:\!\!Reduce(}e_1\x{)}=e_2}{\Ct{e_1} -> \Ct{e_2}} 
\\[15pt]
\inference[reduce($=$)]{}{\Ct{\mathtt{true}} -> \x{i2r(}1\x{)}} 
\end{array}
\]

\subsubsection*{Removal of summations}
  can be done in two ways. Either the index of the summation can only be one value or it can be a limited range of values, and depending on which case  different transformations are used.
  In the first case, there exists an equation containing the variable index of the innermost summation. The equation is solved for the variable, and the rest of the variable occurrences are replaced by the new value.
 \[
 \inference[rem-sum(=)]
{%\x{iso}(\Ct{e_1 \eqa e_2}, \x{x}) -> \x{iso}(\Ct{x \eqa e_3}, \x{x}) & x \notin \mathcal{V}ar(e_3) 
  \x{mm\!\!:\!\!Solve(}e_1 \eqa\! e_2, \x{x}\x{)}\, = \, \Ct{\x{x} \eqa\! e_3}
} 
{\sumq{\x{x}}{\Ct{e_1 \eqa e_2} \tq e} -> 
 e[\x{x} / e_3]}\\ \\
 \]
Removing a summation by its range involves using standard mathematical formulas for rewriting series. The last part of the following rule uses $\sum_{k=1}^{n}k^2 = n(n+1)(2n+1)/6$. 
We only present transformations up to quadratic series 
and our pragmatic implementation
contains rules for transforming series of power of degree up to 10. A more general rewrite rule for series of power of degree up $p$ could be implemented, but is more complicated as it includes Bernoulli numbers and binomial coefficients. The precondition uses Mathematica's \texttt{Expand} to transform the expression into the right pattern. 
 \[
\begin{array}{l}
\inference[rem-sum($\leq$)]
{\!\!\!\mathtt{x}\! \notin\! \mathcal{V}ar(e_1, \ldotss, e_6) \,\,\, \x{mm\!\!:\!\!Expand(}e_3\x{)} = \x{i2r(}e_4 \plusa e_5 \timesa \x{x} \plusa e_6 \timesa \x{x}\timesa \x{x}\x{)}\!\! } 
{ 
\begin{array}{@{}l@{}}
 \sumq{\mathtt{x}}{\Ct{e_1 \leqa \mathtt{x}} \tq \Ct{\mathtt{x} \leqa e_2} \tq \mathtt{i2r}(e_3)} -> \\
\quad \mathtt{i2r}(e_4) \tq \mathtt{i2r}(e_2 \minusa e_1 \plusa 1) \pq \\
\quad \mathtt{i2r}(e_5) \tq \mathtt{i2r}(e_2 \timesa (e_2\! \plusa\! 1))\! \dq 2  \mq \\
\quad \mathtt{i2r}(e_5) \tq \mathtt{i2r}(e_2 \timesa\! (e_2 \minusa\! 1))\! \dq 2 \pq \\
\quad \mathtt{i2r}(e_6) \tq \mathtt{i2r}(e_2 \timesa (e_2\! \plusa\! 1) \timesa  (2 \timesa e_2 \plusa 1)) \! \dq 6 \mq\, \\
\quad \mathtt{i2r}(e_6) \tq \mathtt{i2r}(e_2 \timesa (e_2\! \minusa\! 1) \timesa  (2 \timesa e_2 \minusa 1)) \! \dq 6 
\end{array}
 }   
\end{array}
\]

\subsubsection*{Removal of Product} involves a safe over-approximation. 
The implementation of POA contains two different over-approximations and in 
many cases the probability program can be transformed into closed form in a 
precise manner. In the following paragraph we describe when the transformation 
preserves the accuracy of the transformed term. 

The probability function can always be over-approximated to 1. 
The rule \textsf{f-rec} is an exact rule and introduces a product-expression 
which may not be possible to rewrite into closed form.
We only introduce the product-expression with \texttt{c}-expressions in its 
body, and therefore it will always either evaluate to 1 or 0. 
A safe over-approximation of such a product-expression is 1.
\[\inference[rem-prod-one]
{\mathtt{x} \notin \mathcal{V}ar(e_1,e_2) & \mathtt{x} \in \mathcal{V}ar(e_3)}
{
  \begin{array}{l}
    \prodq{\mathtt{x}}{\Ct{e_1 \leqa \mathtt{x}} \tq \Ct{\mathtt{x} \leqa e_2}}{c(e_3)} -> 1\\
  \end{array}
 }\\ \\ 
\]
For the summation describing recursive calls, this transformation is exact when the condition,  $b$, evaluates to true for exactly one value (eg. it is an equation). 

A broader class of recursive programs (than those having an equation in the condition) 
is those where the \texttt{c}-expression is monotone in $x$; meaning that there 
exists a $k$ for which $c(e_3)=1$ for $x \leq k$ and $c(e_3)=0$ for $x > k$. 
This case covers many for-loops. In this case, we can accurately replace 
the \texttt{prod}-expression with two \texttt{c}-expressions one checking the lower and one checking the upper range-limit. The empty product (the lower limit is larger than the upper) is 1. 
\[\inference[rem-prod-mon]
{\mathtt{x} \notin \mathcal{V}ar(e_1,e_2) & \mathtt{x} \in \mathcal{V}ar(e_3) & e_3\text{ is monotone in }\mathtt{x} }
{
  \begin{array}{l}
    \prodq{\mathtt{x}}{\Ct{e_1 \leqa \mathtt{x}} \tq \Ct{\mathtt{x} \leqa e_2}}{c(e_3)} -> \\
    \big( c(e_3)[\mathtt{x}/e_1] \tq c(e_3)[\mathtt{x}/e_2] \tq \Ct{e_1 \leqa e_2}  \pq \Ct{e_2 \leqa e_1 \minusa 1}\big)
  \end{array}
 } 
\]
This rule does not preserve accuracy when the \texttt{c}-expression is not monotone in $\mathtt{x}$ (e.g. $\Ct{2 \leqa \mathtt{x} || 4 \leqa \mathtt{x}}$).
% If we, in example \ref{ex:app:mon}, had used the \textsf{rem-prod-mon}, then the result would have been accurate: $\mathtt{P}_{\mathtt{f}}(z) = \Ct{z \eqa 3}$. In that example the part to the left of $\pq$ evaluates to 0 and the $\Ct{e_2 \leqa e_1 \minusa 1}$ side forces the \texttt{i}-summation to $\Ct{\x{i}=0}$ leaving only one result. The following example is not monotone in its stop-criteria, and therefore this rewrite rule will result in an over-approximation in that case. 

% ----------------------------------------------------------------------
\section{Implementation and results}

In the following, we present three examples which 
show results of programs with nested loops
parameterized input distribution of multiple variables. The probability distribution computed by the output program varies in complexity; the first program calculates a single parameterized output, the second program computes a triangular shaped output distribution and third computes a distribution converging towards a standard normal distribution.
The results are presented in a reduced and readable form
extracted from our implementation.

\subsection{Matrix multiplication}
The original matrix multiplication program uses composite types and contains nested loops. The intermediate program, defined in Figure \ref{fig:matrix}, contains nested recursive calls but has no dependency on data in composite types. 

\begin{figure}[h!]\centering
\begin{minipage}{\textwidth}\small\begin{verbatim}
 for3(i3,step,n) = if(i3>=n) then step else for3(i3+1,step+1,n)
 for2(i2,step,n) = if(i2>=n) then step else for2(i2+1,for3(0,step+2,n),n)
 for1(i1,step,n) = if(i1>=n) then step else for1(i1+1,for2(0,step,n),n)
 tmulta(step,n) = for1(0,step,n)
 P(step,n1) = c(step=0)*c(n1=n) 
\end{verbatim}
\end{minipage}
\caption{The intermediate program containing also the parameterized probability distribution. The parameter \texttt{n} can obtain only one value.}
\label{fig:matrix}
\end{figure}

The nested calls create argument development functions that depend on function calls. These are transformed into a simple form and then removed. 
The introduced products are over-approximated, but due to the form of the 
condition the result is precise. The output program computes a single value 
distribution (when specialized with the size of the matrix). 
It is given in Figure \ref{fig:matrixRes} along with an array describing a 
subset of specializations of the output program with respect to a value of 
\texttt{n}.
\begin{figure}[h!]
\begin{minipage}{.1\textwidth} 
\begin{tabular}{l}
\end{tabular}
\end{minipage}
\begin{minipage}{.3\textwidth} 
{\small\begin{verbatim}
Ptmulta(out) = 
 c(3=<out/(n*n))*
 c(1=<n)*
 c(out/n*n=2+n)*1
\end{verbatim}}
\end{minipage}
\begin{minipage}{.5\textwidth}\centering
\begin{tabular}{c@{ }|@{ }l} 
 n & program \\ \hline
 1 & {\small\verb+ Ptmulta(out) = c(out=3)+} \\
 2 & {\small\verb+ Ptmulta(out) = c(out=16)+} \\
 3 & {\small\verb+ Ptmulta(out) = c(out=45)+} \\
 4 & {\small\verb+ Ptmulta(out) = c(out=96)+} \\
$\ldots$ & $\qquad\qquad\qquad\,\, \ldots$\\
 \end{tabular}
\end{minipage}
\caption{The general output probability program (left) and the program specialized with the value of \texttt{n} (right).}
\label{fig:matrixRes}
\end{figure}

\subsection{Adding parameterized distributions}

This example is a recursive program computing the addition of two numbers; the input program and the input probability distribution can be seen in Figure \ref{fig:add2}. The output depends on both increasing and decreasing values. In this example, we use a parameter \texttt{n} as the upper limit of a range of input values. The input distribution describes two independent variables, each having a uniform distribution from $1$ to $n$. 

\begin{figure}[h!]
\begin{minipage}{.95\textwidth}\small
\begin{verbatim}
  add(x,y) = if x=<0 then y else add(x-1,y+1)
  P(x) = c(1=<x)*c(x=<n)*1/n
  Pxy(x,y) = P(x)*P(y)
\end{verbatim}
\end{minipage}
\caption{The intermediate program containing both the function \texttt{add} and the input probability distribution. Here, the parameter \texttt{n} is used to describe a range.}
\label{fig:add2}
\end{figure}
\begin{figure}[h!]
\begin{minipage}{.31\textwidth}
\includegraphics[scale=0.5]{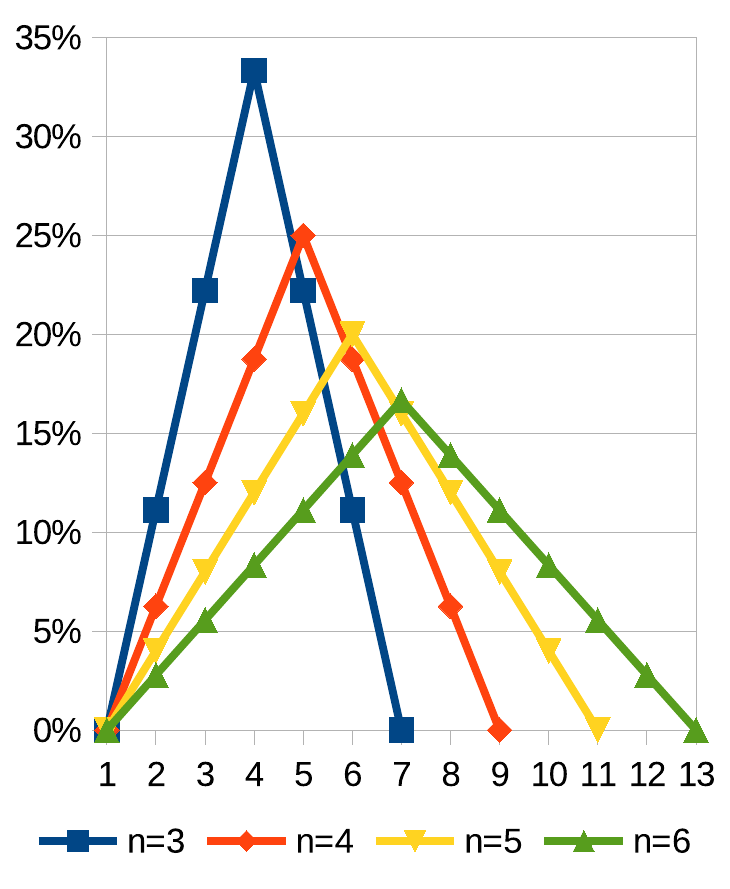}  
\end{minipage}%
\begin{minipage}{.67\textwidth}
\begin{verbatim}
Padd(out) = 
    c(2<=out)*c(out<=n)*(1/n*1/n*(out-1))+
    c(1+n<=out)*c(out<=2*n)*(1/n*1/n*(1+2*n-out))
\end{verbatim}
\end{minipage}
\caption{The general output program and the graphs for the output probability distribution with n set to 3, 4, 5, and 6, respectively.}
\label{fig:add2Res}
\end{figure}

  The analysis gives a precise probability distribution and computes a triangular distribution (or pyramid shaped distribution). The output probability program is shown in Figure \ref{fig:add2Res} along with a graph depicting the pyramid shaped output probability distributions for different initializations of \texttt{n}.
The lower bound on \texttt{out} arises from the input probability distribution and not from the condition. 
The upper bound \verb+2*n+ of the analysis result shows that 
%transformation can deduct that 
the output depends on both input variables, despite that one is increasing and the other is decreasing.

\subsection{Adding 4 independent variables}
The program \texttt{sum4} adds four variables and was
presented by Monniaux \cite{conf/sas/Monniaux00}. 
Certain over-approximations were applied so as to obtain a safe and simplified result.
% who analyzed programs with real distributions as input. He uses abstract intepretation to over approximate real distributions and afterwards execute the program. 
%The result was that summed the over aproximated numbers, instead of, as here, an over approximation of the sum of precise numbers. 

%The original program and the step-counting program are similar as the number of steps depends directly on the size of all four variables. 
The  program is recursive and in this example 
we use independent input variables each uniformly distributed input from 1 to 6, as described in Figure \ref{fig:add4}.

\begin{figure}[h!]
\begin{minipage}{.95\textwidth}\small
\begin{verbatim}
   add(x,y) = if x=0 then y else add(x-1,y+1)
   sum4(x,y,z,w) = add(x,add(y,add(z,w)))
   tsum4(x,y,z,w) = sum4(x,y,z,w)
   P(x) = c(1=<x)*c(x=<6)*1/6
   Pxyzw(x,y,z,w) = P(x)*P(y)*P(z)*P(w)
\end{verbatim}
\end{minipage}
\caption{Intermediate program.}
\label{fig:add4}
\end{figure}

\begin{figure}[h!]
\begin{minipage}{.51\textwidth}
\includegraphics[scale=0.5]{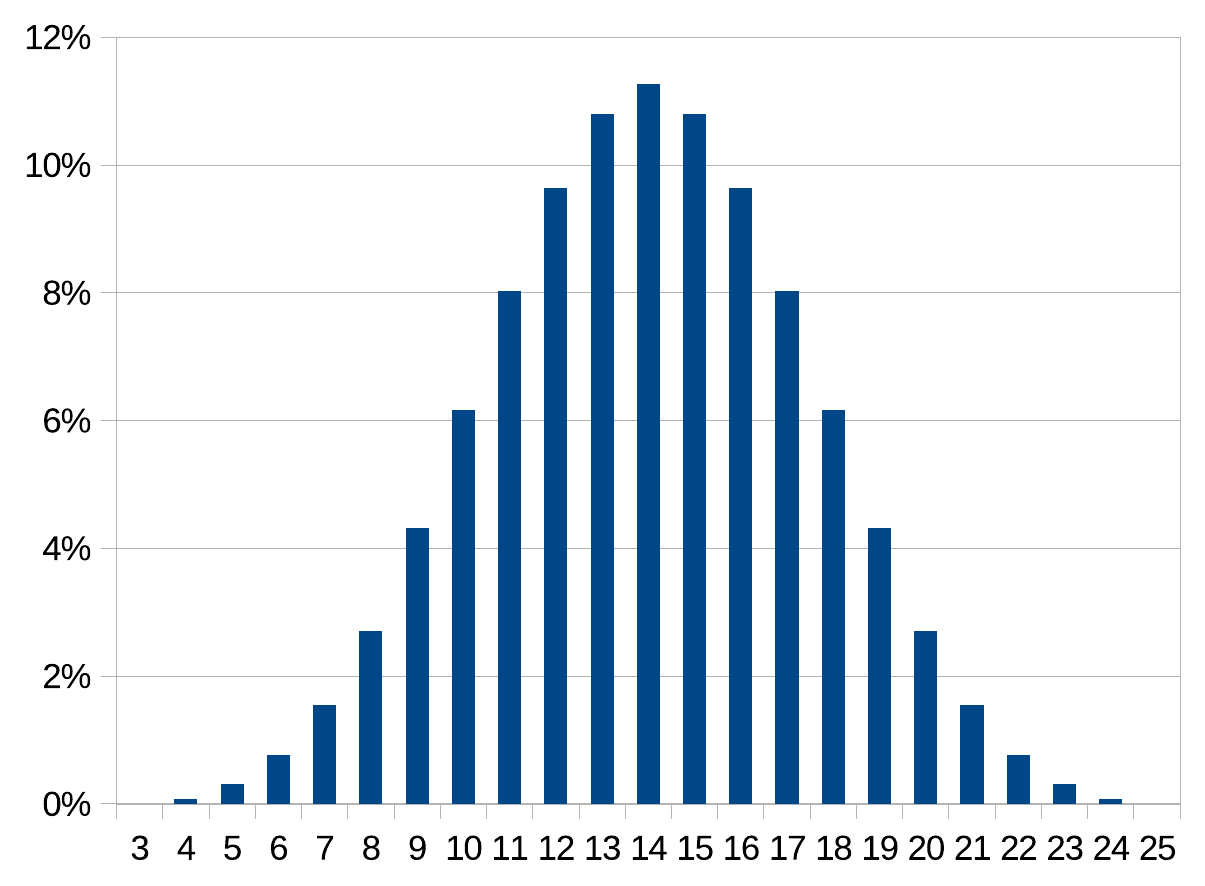}  
\end{minipage}%
\begin{minipage}{.49\textwidth}\small
\begin{verbatim}
Psum4(out) =
c(4=<out)*c(out=<7)*(-6 + 11*out - 
6*out^2 + out^3)/7776+
c(8=<out)*c(out=<12)*
(-1014+169*out+6*out^2-out^3)/7776+
c(9=<out)*c(out=<12)*
(1512-461*out+42*out^2-out^3)/3888+
c(out=13)*(265/648-5*out/216)+
c(14=<out)*c(out=<18)*
(-4790+923*out-54*out^2+out^3)/2592+
c(19=<out)*c(out=<24)*
(17550-2027*out+78*out^2-out^3)/7776
\end{verbatim}
\end{minipage}
\caption{The output program and graph for its computed probability distribution for \texttt{out} from 3 to 25.} 
\label{ref:add4Res}
\end{figure}
  Despite the ranges and their associated value are not symmetric, the resulting program computes a precise and perfectly symmetric probability distribution as shown in Figure \ref{ref:add4Res}. 
The difference in the ranges comes (among other things) from the range dividing rules, as they do not divide the range symmetrically. 
As expected from the central limit theorem of probability theory, the resulting probability program describes a distribution that has similarities with a normal distribution. 

\subsection{Monty Hall}
The Monty Hall problem is often used to exemplify how gained knowledge influences 
probabilities (conditional probability). 
In this problem, there are three closed doors; one hiding a price and two 
that are empty. The doors have an equal chance of hiding the price. 
There is a contestant, who should choose one of the doors, then the game 
host will open an empty door and the contestant can either stick with the 
first choice or can change to the other unopened door. 
The problem lies in showing whether the best winning-strategy is to 
stick with the first choice or to switch to the other? 

If the strategy is to stick with the first choice and that door has a price 
then the contestant has won. If the contestant changes door he/she only loses 
if the first choice was the door hiding the price; if the first choice was an 
empty door, then the game host would open the other empty door leaving only the 
price door for a second choice.

The program \texttt{monty} models the two strategies; if the strategy variable 
is 1 then the strategy is to change the door, and otherwise the strategy is to 
stick with the first choice. The program takes as input the contestant's first 
guess, the door hiding the price, the empty door which is not opened by the 
game host and the strategy the contestant uses. 

Let us assume the contestant has an equal chance of choosing each of the doors. The input variables \texttt{guess}, \texttt{price}, and \texttt{empty} models the first choice, the price door and the empty door which is left after the game host has opened an empty door. All three doors have a value between 1 and 3, and the empty door cannot be the same as the price door. We have parameterized the strategy with a weight \texttt{p} between the two, such that when \texttt{p} = 1 then the strategy is to always change door, and when \texttt{p}=0 the strategy is to always keep the first choice (e.g. letting \texttt{p} = 0.75 we change doors in 3/4 cases and 1/4 we keep the first door). Such a parameterization allows us to execute the analysis once and use the lighter closed form result for that calculation instead. In a problem where the winning-probability of a strategy is dependent on the other input, such input could be used for optimizing the choice of strategy. The program \texttt{monty} and the parameterized input probability distribution can be seen in Figure \ref{fig:monty}. %Then we can describe the doors using relations; there are three doors: one with a price, an empty door that the gamehost will open that can be neither the price door nor the chosen door and the other empty door which can be neither the price door nor the empty door opened by the game host. 

\begin{figure}[h!]
\hspace{0.02\textwidth}%
\begin{minipage}{.48\textwidth}\small
\begin{verbatim}
monty(guess,price,empty,strategy)= 
  if strategy = 0 
  then finalGuess(guess,price) 
  else change(guess,price,empty)

finalGuess(guess,price)= 
  if price=guess then 1 else 0

change(guess,price,empty)= 
  if price=guess 
  then finalGuess(empty,price) 
  else finalGuess(price,price)
\end{verbatim}
 \end{minipage}%
 \begin{minipage}{.48\textwidth}\small
\begin{verbatim}
Pin(guess,price,empty,strategy) =  
  1/18*c(1=<guess)*c(guess=<3) 
      *c(1=<price)*c(price=<3)
      *c(1=<empty)*c(empty=<3)
      *c(not(price = empty))
      *Pstrat(strategy)
Pstrat(strategy) = 
  p*c(1 = strategy) 
    + (1-p)*c(0 = strategy)  
\end{verbatim}
 \end{minipage}
\caption{The program \texttt{monty} models the event flow depending on the chosen strategy; if the strategy is \texttt{0} then the contestant keeps the first door and if it is \texttt{1} then the contestant changes his mind. 
There are three doors and the input of \texttt{monty} describes the contestants first guess, the door hiding the price, the empty door which is not opened by the game host (and is different from the price door) and the strategy of the contestant. If the final choice hides the price then the program returns 1 and otherwise 0.
The probability of the strategy is an expression parameterized with a weight, \texttt{p} between the two strategies instead of executing the analysis twice.
}
\label{fig:monty}
\end{figure}

\begin{figure}[h!]
 \begin{minipage}{.32\textwidth}\small
  \begin{verbatim}
pmonty(out) =  
 1/18 * 
 (c(out=0)*
   (12*(1-p)+6*p)+ 
  c(out=1)*
   (6*(1-p)+12*p))
 \end{verbatim}
 \end{minipage}
 \begin{minipage}{.67\textwidth}  
\includegraphics[scale=0.5]{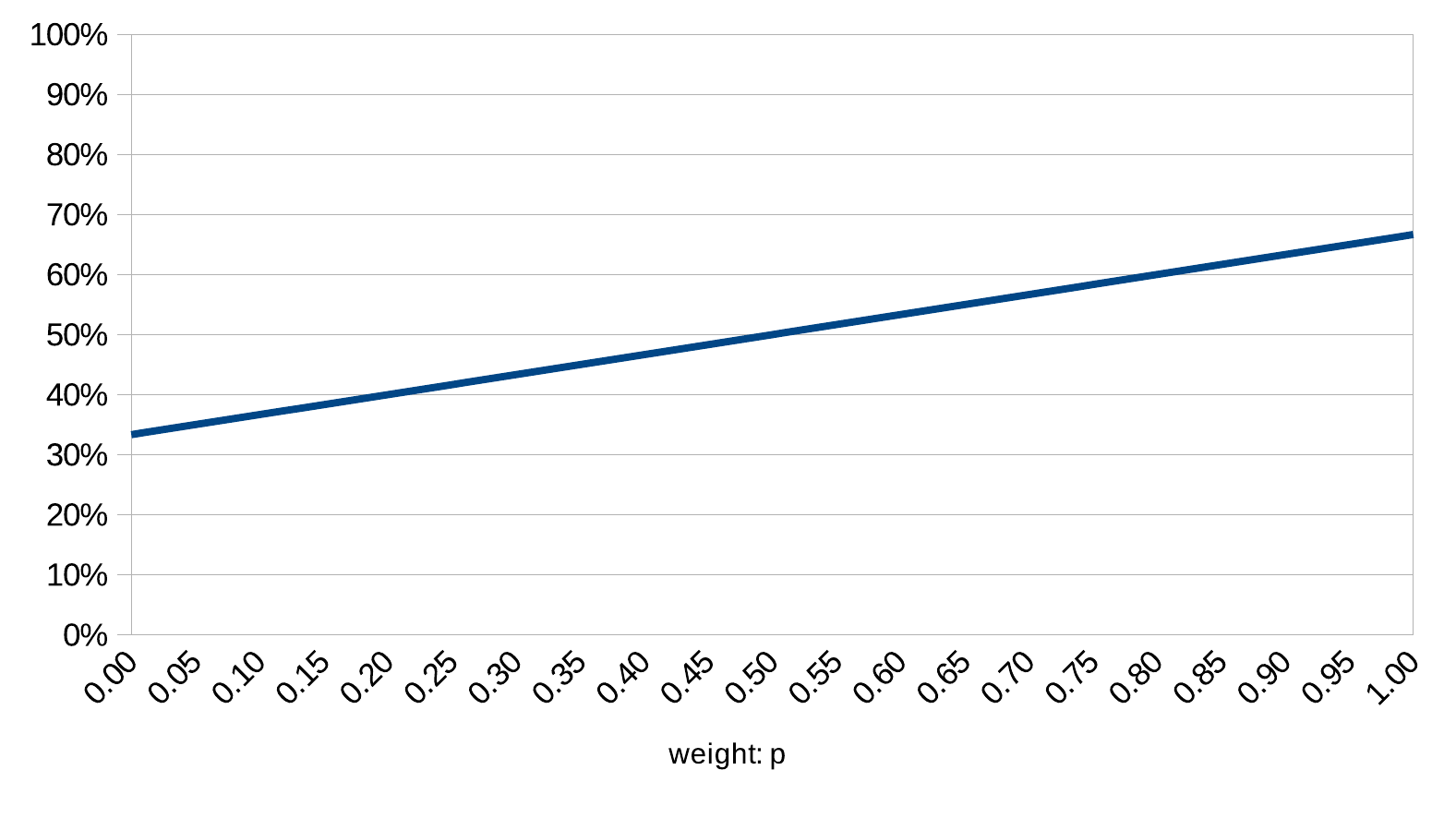} 
 \end{minipage}
\caption{The probability of winning the Monty Hall as a function of the weight 
given to change-strategy.
The probabilistic output analysis reveals that the best weight between the keep 
strategy and the change strategy is to always use change strategy.}
\label{fig:montyRes}
\end{figure}

The analysis was capable of handling the program correctly and the result can 
be seen in Figure \ref{fig:montyRes}. 

The probabilities 1/3 and 2/3 does not occur directly in the output probability 
program but are found in the constants 6, 12 and 1/18.

\subsection{Adding dependent non-uniform variables}
A function call may have interdependent and non-uniform arguments, 
and in this example, we demonstrate that the analysis can handle such function calls. 
We focus on the dependencies, analyze a simple add program and discuss the 
limits of the interdependencies. The program also shows that interdependencies 
quickly lead to the occurrence of integer division in the output 

The input arguments are interdependent; the second argument is always less than or equal to the value of the first argument. The joint distribution depends only on the value of the first argument resulting in a skewed probability distribution. The probability program is defined in Figure \ref{fig:addDep}.

\begin{figure}[h!]
\begin{minipage}{\textwidth}\small
\begin{verbatim}
  Pxy(x,y) =  c(1=<y)*c(y=<3) * 
               c(1=<x)*c(x=<y) * x/10
  add(x,y) = x+z

  Padd(out) = 
    c(2 =< out)*c(out=< 3) *   1/20 * out%2 * (1 + out%2) +
    c(4 =< out)*c(out=< 6) * -(1/20)*(-4+out-out%2)*(-3+out+out%2)
\end{verbatim}
\end{minipage}
\caption{An input program, \texttt{add}, its skewed joint distribution, \texttt{Pxy}, and the closed form probability program, \texttt{Padd}, produced by the analysis. The integer division is noted by a ``\texttt{\%}''.}
\label{fig:addDep}
\end{figure}

The create rule generates nested summations, and removing such inner summations imply that their values must be expressed using the variables of the outer summations or the input variable (ie. \texttt{out}). Comparing the result from this experiment with the output probability distribution for addition of two random variables in Figure \ref{fig:add2Res} indicates that integer division is a special case arising from dependent input. The following interesting expressions are extracted during analysis execution, and they shows how the integer division arises from the dependency of input. The first expressions is the result from the create rule and the last expression is the result after removal of the inner $\x{y}$-summation.

{\footnotesize
\[ 
\begin{array}{l}
P_{\mathtt{add}}(\x{out}) = \\
\sumq{\x{x}}{\sumq{\x{y}}{ \Ct{\x{out}  \eqa \x{x} \plusa \x{y}} \tq\\
\qquad   \Ct{1 \leqa \x{x}} \tq \Ct{\x{x} \leqa \x{y}} \tq \Ct{1 \leqa \x{y}} \tq 
  \Ct{\x{y} \leqa 3} \tq 
   ( \bnfts{i2r(}\x{x}\bnfts{)} \dq 
   \bnfts{i2r(}10\bnfts{)}) }} = \\\\ 
  \sumq{\x{x}}{  \Ct{2 \leqa \x{out}} \tq \Ct{\x{out} \leqa 3} \tq  
    \Ct{1 \leqa \x{x}} \tq 
   \\\;\;\;
    \Ct{2 \timesa \x{x} \leqa \x{out}} \tq 
    ( \bnfts{i2r(}\x{x}\bnfts{)} \dq \bnfts{i2r(}10\bnfts{)})} 
\pq \\
 \sumq{\x{x}}{ \Ct{4 \leqa \x{out}} \tq \Ct{\x{out} \leqa 3 \plusa \x{x}} \tq 
   \Ct{2 \timesa \x{x} \leqa \x{out}} \tq 
     ( \bnfts{i2r(}\x{x}\bnfts{)} \dq \bnfts{i2r(}10\bnfts{)}) }
\end{array}
\]}

In the last expression, there are two summations, each leading to its own part in the resulting program. Looking closely at each summation, we see that they share the upper limit for $\x{x}$, $\Ct{2 \timesa \x{x} \leqa \x{out}}$, which currently contains an integer multiplication and when solved with respect to $\x{x}$ contains the integer division. In the final result, the second part of the expression has an upper limit for \texttt{out}, $\Ct{\x{out} \leqa 6}$ which is a constraint that the summation-removal-rule introduces to ensure that the lower limit of the summation (i.e. $\x{out} \minusa 3$) is less than or equal to the upper limit (i.e. $\x{out} \diva 2$).

The original probability $(\bnfts{i2r(}\x{x}\bnfts{)} \dq \bnfts{i2r(}10\bnfts{)})$ 
occurs directly in the summations, and this indicates a limit of this 
implementation and approach. To be able to handle a probability, the rewrite 
rules for summations must transform summations over the probability expression. 
There are limits to which series the system currently can transform, 
Sum of
reciprocals (e.g. $\sum_{k=1}^n \frac{1}{k}$) known as harmonic series or 
variations thereof such as generalized harmonic series are currently not implemented. 
The current analysis is 
limited to finite summations of at least order of 1, but a closer integration 
with Mathematica that exploits more of Mathematicas rewriting mechanisms should be 
able to handle such series.

%-----------------------------------------------------
\section{Related works}
\label{sec:relatedWork}

Probabilistic analysis is related to the analysis of probabilistic programs. 
Probabilistic analysis analyze programs with a normal semantics where the input variables are interpreted over probability distributions. 
Analysis of probabilistic programs analyze programs 
with probabilistic semantics where the values of the input variables 
are unknown (e.g. flow analysis \cite{conf:pfis:Pierro2013}).

In probabilistic analysis, it is important to determine how variables depend on each other, but already in 1976 Denning proposed a flow analysis for revealing whether variables depend on each other \cite{journals/cacm/Denning76}. 
This was presented in the field of secure flow analysis.
Denning introduced a lattice-based analysis where she, given the name of a variable, that should be kept secret, deducted which other variables those should be kept secret in order to avoid leaking information. 
In 1996, Denning's method was refined by Volpano {\etal} into a type system and for the first time, it was proven sound \cite{journals/jcs/VolpanoIS96}. 

Reasoning about probabilistic semantics is a closely related area to probabilistic analysis, as they both work with nested probabilistic influence. 
The probabilistic analysis work on standard semantic and analyze it using input probability distributions, where a probabilistic semantics allow for random assignments and probabilistic choices 
\cite{journals/jcss/Kozen81} 
%\cite{conf/focs/Kozen79} 
and is normally analyzed using an expanded classical analysis or verification method \cite{conf/esop/CousotM12}.

Probabilistic model checking is an automated technique for formally verifying quantitative properties for systems with probabilistic behaviors. 
It is mainly focused on Markov decision processes, which can model both stochastic and non-deterministic behavior 
\cite{conf/sfm/ForejtKNP11,conf:alleton:Kwiatkowska2010}.
It differs from probabilistic analysis as it assumes the Markov property.

In 2000, Monniaux applied abstract interpretation to programs with probabilistic semantics and gained safe bounds for worst-case analysis 
\cite{conf/sas/Monniaux00}.
Pierro {\etal} introduce a linear mapping structure, a Moore-Penrose pseudo-inverse, instead of a Galois connection. 
They use the linear structures to compare 'closeness' of approximations as an expression using the average approximation error.  
Pierro {\etal} further explores using probabilistic abstract interpretation to calculate the average-case analysis 
\cite{conf/birthday/PierroHW06}.
In 2012, Cousot and Monerau gave a general probabilistic abstraction framework
\cite{conf/esop/CousotM12}
and stated, in section 5.3, that Pierro {\etal}'s method and many other abstraction methods can be expressed in this new framework.

When analyzing probabilities the main challenge is to maintain the dependencies throughout the program. 
Schellekens defines this as \textit{Randomness preservation} \cite{books/daglib/0020847} (or random bag preservation) which in his (and Gao's \cite{thesis:cork:gao2013}) case enables tracking of certain data structures and their distributions. 
They use special data structures as they find these suitable to derive the average number of basic operations. 
In another approach \cite{journals/cacm/Wegbreit75,journals:ing:PollmanCG09}, tests in programs has been assumed to be independent of previous history, also known as the Markov property (the probability of true is fixed). 
As Wegbreit remarked, this is true only for some programs (e.g. 
linear search for repeating lists) and others, this is not the case (linear search for non-repeating lists). 
The Markov property is the foundation in Markov decision processes which is used in probabilistic model-checking \cite{conf/sfm/ForejtKNP11}. 
Cousot et al. presents a probabilistic abstraction framework where they divide the program semantics into probabilistic behavior and (non-)deterministic behavior. 
They propose handling of tests when it is possible to assume the Markov property, and handle loops by using a probability distribution describing the probability of entering the loop in the $i$th iteration. 
Monniaux proposes another approach for abstracting probabilistic semantics \cite{conf/sas/Monniaux00}; he first lifts a normal semantics to a probabilistic semantics where random generators are allowed and then uses an abstraction to reach a closed form. 
Monniaux's semantic approach uses a backward probabilistic semantics operating on measurable functions. 
This is closely related to the forward probabilistic semantics proposed earlier by Kozen~
%\cite{conf/focs/Kozen79}
\cite{journals/jcss/Kozen81}. 
\nocite{conf:lopstr:Liqat2013}
\nocite{msc:ku:Rosendahl1986}

An alternative approach to probabilistic analysis is based on symbolic execution of programs with symbolic values
\cite{conf:issta:Geldenhuys2012}. 
Such techniques can also be used on programs with infinitely many execution paths 
by limiting the analysis to a finite set of paths at the expense of 
tightness of probability intervals \cite{conf:pldi:SankaranarayananC2013}.

%-----------------------------------------------------
\section{Conclusion}

Probabilistic analysis of program has a renewed interest for analyzing programs for energy consumptions.
Numerous embedded systems and mobile applications are limited by restricted battery life on the
hardware. In this paper, we describe a rewrite system that derives a resource probability distribution for programs given distributions of the input. 
It can analyze programs in subset of C where we have known distribution of input variables. 
From the original program we create a probability
distribution program, where we remove calls to original functions and 
transform it into closed form. 
We have presented the transformation rules for each step and outlined the implementation of the system. 
We discuss over-approximating rules and their influence on the accuracy of the output probability and show that our analysis improves on related analysis in the literature.
%The presented over-approximation rules is shown to 
%improve the output analysis by Monniaux \cite{conf/sas/Monniaux00}.

\nocite{zuleger2011bound}
\bibliographystyle{splncs}
\bibliography{dir-ref}

\end{document}